# Numerical Study of Two-fluid Flowing Equilibria of Helicity-driven Spherical Torus Plasmas

T. Kanki[*], M. Nagata[†], and T. Uyama[†]

[*]*Japan Coast Guard Academy, 5-1 Wakaba, Kure, Hiroshima 737-8512, Japan*
[†]*Department of Electrical Engineering and Computer Sciences, University of Hyogo,
2167 Shosha, Himeji, Hyogo 671-2201, Japan*

**Abstract.** Two-fluid flowing equilibrium configurations of a helicity-driven spherical torus (HD-ST) are numerically determined by using the combination of the finite difference and the boundary element methods. It is found from the numerical results that electron fluids near the central conductor are tied to an external toroidal field and ion fluids are not. The magnetic configurations change from the high-$q$ HD-ST ($q>1$) with paramagnetic toroidal field and low-$\beta$ (volume average $\beta$ value, $<\beta> \approx 2$ %) through the helicity-driven spheromak and RFP to the ultra low-$q$ HD-ST ($0<q<1$) with diamagnetic toroidal field and high-$\beta$ ($<\beta> \approx 18$ %) as the external toroidal field at the inner edge regions decreases and reverses the sign. The two-fluid effects are more significant in this equilibrium transition when the ion diamagnetic drift is dominant in the flowing two-fluid.

## INTRODUCTION

During coaxial helicity injection, toroidal ion flow related to a rotating toroidal mode number $n=1$ magnetic structure has been observed in many helcity-driven spherical torus (HD-ST) experiments such as CTX, HIST, SPHEX, SSPX, HIT, and NSTX. The $n=1$ mode structure is considered to be playing an essential role in driving current on closed flux surfaces of the HD-ST. On the HIT experiments [1], it is found that the $n=1$ mode is locked to electrons and not to ions, suggesting a rotating magnetic field current drive. Because of this feature, the equilibrium computation of a HD-ST is required to take into account two-fluid effects [2-3]. The two-fluid effects are expected to explain the stability of high-$\beta$ ST. However, the details of how such the flowing two-fluid model affects the MHD equilibrium configurations of the HD-ST are not numerically investigated.

The purpose of this study is to numerically determine the two-fluid flowing equilibria of the HD-ST and to investigate their fundamental properties. We focus our attention on contribution of the ion flow to the magnetic configuration, the two-fluid effects, and $\beta$ values. The formalism for flowing two-fluid equilibrium is developed by Steinhauer, Ishida and co-workers [4]. It is an extension of the MHD equilibrium problem for a non-flowing single-fluid which is governed by the Grad-Shafranov equation. The axisymmetric equilibrium of the flowing two-fluid is described by a pair of second-order partial differential equations for the magnetic and ion flow stream functions, and Bernoulli equation for the density [4]. By applying the two-fluid formulation to the HD-ST equilibrium with purely toroidal ion flow, we modify the non-flowing single-fluid equilibrium code [5] which computes the HD-ST equilibrium in the more realistic region including the spherical flux conserver (FC) and the coaxial helicity source (CHS) of HIST [6]. In this code, the equilibrium computation of the HD-ST reflects the realistic condition that the bias coil flux penetrates the FC wall and the electrodes. In order to solve the governing equations of the flowing two-fluid equilibrium, we employ the finite difference and the boundary element methods as the numerical approach incorporating this boundary condition.

## NUMERICAL MODEL

For numerical computation, we model the more realistic region including the FC and the CHS of HIST. According to HIST geometry, the spherical FC is 1.0 m in diameter. The FC adjoins the CHS. The CHS consists of the outer electrode (0.276 m in diameter, 0.35 m in length), the inner electrode (0.18 m in diameter, 0.309 m in length), and the outer bias coil (0.362 m in diameter, 0.3 m in length) of rectangular cross section. The central conductor (0.114 m in diameter) is inserted along the symmetry axis. Insertion of a toroidal field coil current $I_{tf}$ along the geometry axis inside the central conductor produces an external toroidal field. In Fig. 1 we show the model of the FC and the CHS which will be used in this paper. We divide the region in which the equilibrium is determined into three subregions, $\Omega_1$, $\Omega_2$, and $\Omega_3$. In the HIST experiment, the bias field is generated long before the plasma is injected into the FC. Thus, the bias field penetrates the FC wall, the electrodes, and the central conductor, and extends all over the space when the equilibrium configuration is formed. On the other hands, the lifetime of the plasma is much shorter than the resisitive penetration time of the FC, the inner electrode, and the central conductor and it is much longer than that of the outer electrode. Therefore, we assume that the magnetic field generated by the plasma current penetrates the outer electrode and that it does not penetrate the FC wall, the inner electrode and the central conductor.

Let us use a cylindrical coordinate system $(r, \theta, z)$ in which the $z$-axis lies along the symmetry axis of HIST geometry. Since the two-fluid flowing equilibrium configuration of the HD-ST plasma is axially symmetric, we can determine it by solving the coupled pair of differential equations for the generalized stream functions $\Psi_i$ and $\Psi_e$. The coupled equations can be written in the form [4],

$$r^2 \frac{d\psi_i}{d\Psi_i} \nabla \cdot \left( \frac{d\psi_i}{d\Psi_i} \frac{\nabla \Psi_i}{r^2} \right) = S_*^2 (\psi_i - \psi_e) \frac{d\psi_i}{d\Psi_i} + S_*^2 (\Psi_e - \Psi_i) + r^2 \frac{dH_i}{d\Psi_i}, \quad (1)$$

$$\Delta^* \Psi_e = S_*^2 (\psi_i - \psi_e) \frac{d\psi_e}{d\Psi_e} + S_*^2 (\Psi_e - \Psi_i) - r^2 \frac{dH_e}{d\Psi_e}. \quad (2)$$

Here $\Delta^*$ denotes the Grad-Shafranov operator and $S_*$ is defined as the ratio of the system size scale to the ion skin depth. The flow stream functions $\psi_\alpha$ and the generalized stream functions $\Psi_\alpha$ are introduced to express the species flow velocities and the poloidal part of the generalized vorticitiy of each species, respectively. The total enthalpies $H_\alpha$ and $\psi_\alpha$ are arbitrary surface functions of their respective surface variables $\Psi_\alpha$, respectively,

$$H_\alpha(\Psi_\alpha) = p_\alpha + u_\alpha^2 / 2 + q_\alpha \phi_E, \quad (3)$$
$$\psi_\alpha = \psi_\alpha(\Psi_\alpha). \quad (4)$$

Here $p_\alpha, u_\alpha, q_\alpha,$ and $\phi_E$ are the pressure, the flow velocity, charge, and the scalar potential regarding the steady electric field. In this study we consider the special case of purely toroidal ion flow, i.e., $\psi_i(\Psi_i) = 0$. In addition, we assume the remaining arbitrary functions:

$$dH_i / d\Psi_i = C_{Hi0} + C_{Hi1} \Psi_i, \quad (5)$$
$$dH_e / d\Psi_e = C_{He0} + C_{He1} \Psi_e + C_{He3} \Psi_e^3, \quad (6)$$
$$\psi_e(\Psi_e) = C_{BT} + C_{\psi e1} \Psi_e + C_{\psi e2} \Psi_e^2, \quad (7)$$

where $C$'s are constant parameters. Especially, $C_{Hi1}$ and $C_{BT}$ are related to the strength of ion flow and external toroidal field, respectively. We change these two parameters. Next let us consider the boundary conditions for Eqs. (1) and (2). We set $\Psi_e = \Psi_{bias}$ on $\Gamma_1$, $\Gamma_3$, $\Gamma_4$ and $\Gamma_5$ because the bias flux extends all over the space. Here $\Psi_{bias}$ represents the bias flux produced by the bias coil current $I_{bias}$. Ampere's law on the surface $\Gamma_c$ can be written as

$$I_{bias} = \oint_{\Gamma_c} \frac{1}{r} \frac{\partial \Psi_e}{\partial n} dl. \quad (8)$$

Here $\mathbf{n}$ denotes the unit vector whose direction is outward normal to the boundary. The boundary condition for $\Psi_e$ on the surface $\Gamma_c$ is obtained by setting at an unknown constant. After the linearization of Eqs. (1) and (2), the problem can be solved numerically by means of the combination of the finite difference and the boundary element methods [5]. This computation is performed so that the total toroidal current $I_t$ is constant.

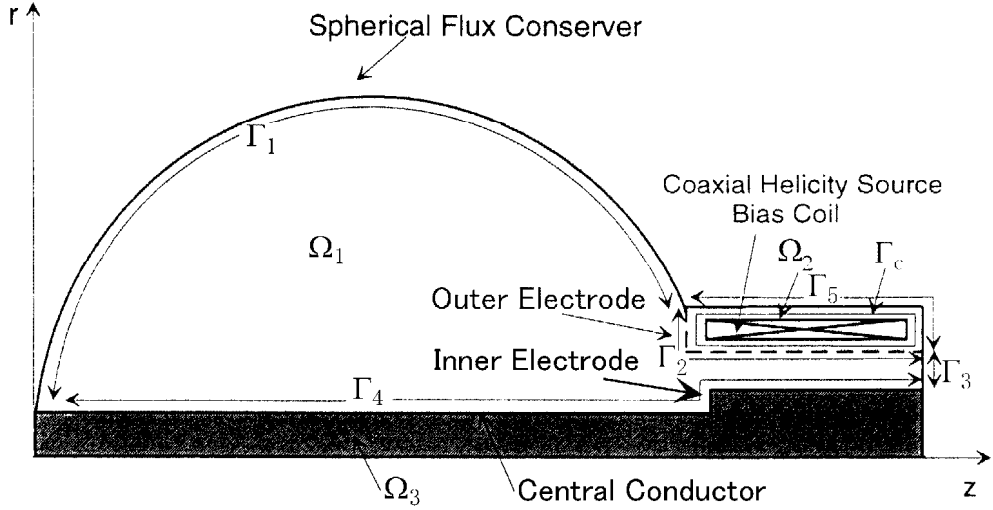

**FIGURE 1.** Model of the flux conserver and the coaxial helicity source.

## NUMERICAL RESULTS

We increase the value of $C_{Hi1}$ related to the strength of ion flow to investigate the variation of the magnetic configuration. As the result of iterations, the value of $S_* C_{BT}$ is determined as the eigenvalue. Parameters, and computed various values such as volume average $\beta$ value $<\beta>$, volume average toroidal $\beta$ value $<\beta_T>$, normalized $\beta$ value $\beta_N$, two-fluid index $f_{2F}$, and volume average $\lambda$ value $<\lambda>$ are shown in Table I. Here $<\beta>$, $<\beta_T>$, $f_{2F}$ and $<\lambda>$ are defined as

$$<\beta> \equiv \frac{<p_i + p_e>}{<p_i + p_e + B^2/2\mu_0>}, \qquad (9)$$

$$<\beta_T> \equiv \frac{<p_i + p_e>}{B_{t0}^2/2\mu_0}, \qquad (10)$$

$$f_{2F} \equiv \frac{<|\mathbf{F}_{2F} \times \mathbf{B}|>}{<|\mathbf{E} \times \mathbf{B}|>}, \qquad (11)$$

$$<\lambda> \equiv \mu_0 \frac{<\mathbf{j} \cdot \mathbf{B}>}{<B^2>}. \qquad (12)$$

Here $\mathbf{B}$, $\mathbf{j}$, $\mathbf{E}$, and $\mathbf{F}_{2F}$ represent magnetic field, current density, electric field, and two-fluid correction term in Ohm's law, $\mathbf{E} + S_* \mathbf{u}_i \times \mathbf{B} + \mathbf{F}_{2F} = 0$, respectively. Also, $B_{t0}$ is the vacuum toroidal field at the major radius $R_0$. The average is over the entire region of $\Omega_1$. If $f_{2F} \geq 1$ ($f_{2F} << 1$), the two-fluid effect is significant (negligible).

The magnetic field profiles on the midplane are shown in Fig. 2. Figure 2(a) shows the case of small ion flow ($C_{Hi1}=1.0$) and the high-$q$ ST with paramagnetic toroidal field $B_t$ profile. We indicate the flow velocity **u** and the safety factor $q$ later. As the effect of the ion flow becomes more significant, the external toroidal field $B_{t,e}$ decreases and further reverses its sign. Figure 2(b) shows the spheromak configuration without $B_{t,e}$. In Fig. 2(c), $B_t$ at the edge regions reverses the sign, which indicates the RFP-like configuration. Eventually, $B_t$ at the whole regions reverses the sign as shown in Fig. 2(d). The magnetic configuration then changes to the ultra low-$q$ ST with diamagnetic $B_t$ profile.

**TABLE I.** Parameters, and computed various values of helicity-driven spherical torus equilibria for $C_{Hi0}=0.0$, $C_{He0}=0.0$, $C_{He1}=4.0$, $C_{He3}=-1.0$, $S_* C_{\psi e1}=-0.7$, $S_* C_{\psi e2}=0.0$, and $I_{bias}/I_t=2.0$.

| $C_{Hi1}$ | $S_* C_{BT}$ | $\langle\beta\rangle$ | $\langle\beta_T\rangle$ | $\beta_N$ | $f_{2F}$ | $\langle\lambda\rangle$ [m$^{-1}$] |
|---|---|---|---|---|---|---|
| 1.0 | -0.240 | $2.30\times10^{-2}$ | $8.11\times10^{-2}$ | 1.55 | 1.83 | 1.32 |
| 9.2 | $-5.02\times10^{-3}$ | 0.146 | ----------- | ----- | 2.82 | 2.22 |
| 15.0 | 0.215 | 0.230 | ----------- | ----- | 1.87 | 2.24 |
| 28.0 | 0.926 | 0.179 | 0.135 | 11.3 | 1.50 | -0.673 |

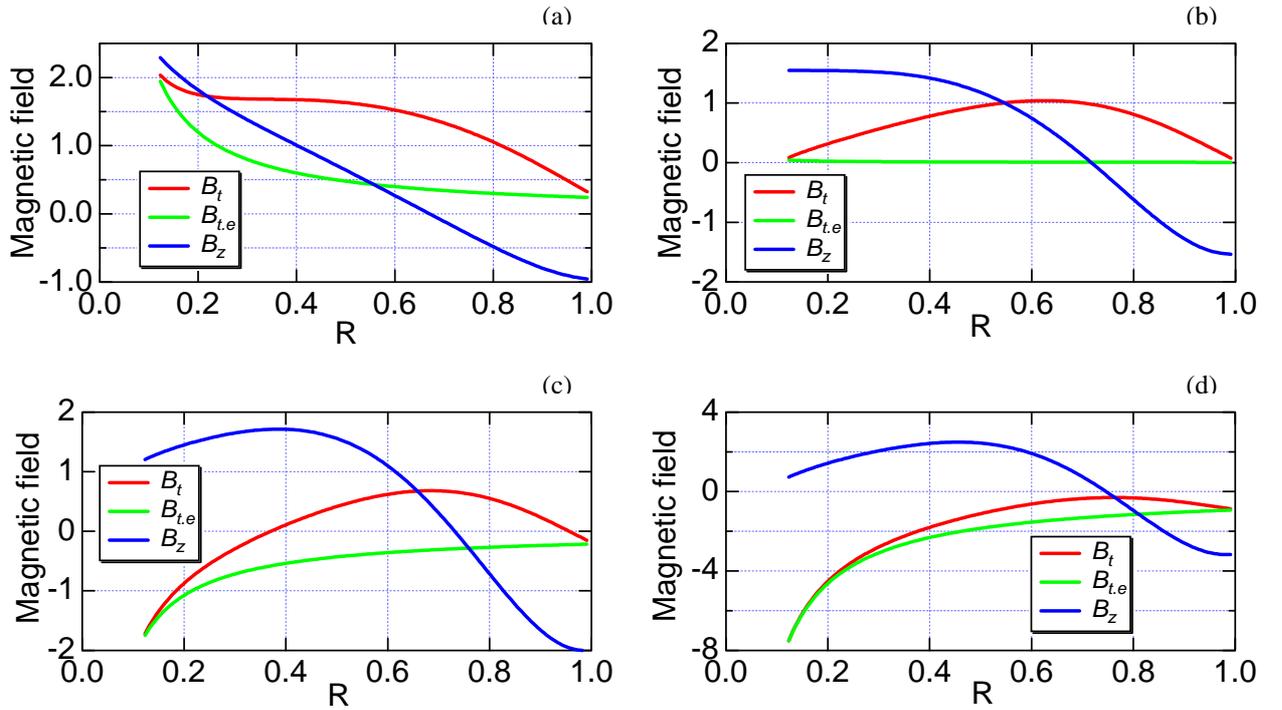

**FIGURE 2.** Radial profiles of magnetic field on the midplane. (a) high-$q$ HD-ST ($C_{Hi1}=1.0$), (b) helicity-driven spheromak ($C_{Hi1}=9.2$), (c) helicity-driven RFP ($C_{Hi1}=15.0$), and (d) ultra low-$q$ HD-ST ($C_{Hi1}=28.0$). The red, green, and blue lines indicate the toroidal field, the external toroidal field, and the poloidal field, respectively.

The flow velocity profiles on the midplane are shown in Fig. 3. It is found from Fig. 3(a) that the toroidal current is dominantly carried by the electron fluids. The electron fluids at the inner edge region are tied to $B_{t,e}$ while the ion fluids are not. Figure 3(c) shows the electron flow at the inner edge region reverses the sign due to the reversal of $B_{t,e}$ at the inner edge region. As the effect of the ion flow becomes more significant, the reversed region of the toroidal electron flow extends as shown in Fig. 3(d).

The toroidal current density profiles on the midplane are shown in Fig. 4. As the effect of the ion flow becomes larger, the toroidal current density changes from the hollow profile to the peaked one. Further, due to the reversal of the toroidal electron flow, it reverses the sign at the inner edge region.

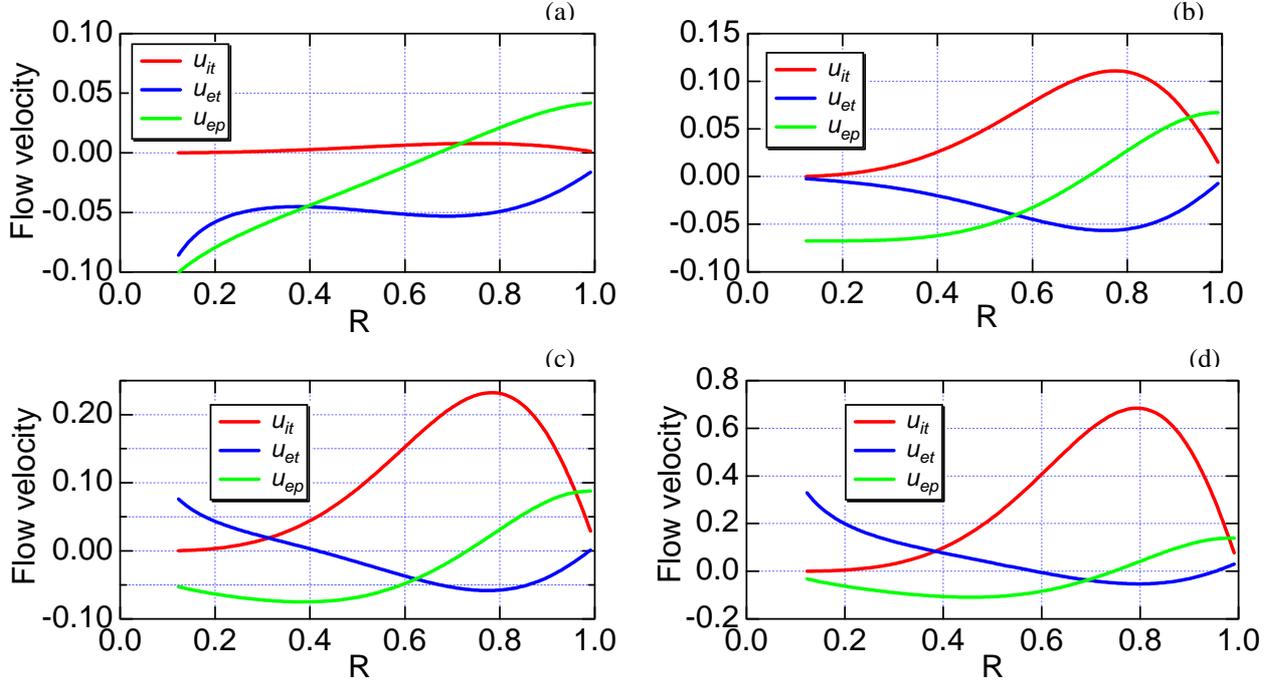

**FIGURE 3.** Radial profiles of flow velocity on the midplane for the same condition as Fig. 2. The red, blue, and green lines indicate the ion toroidal flow, electron toroidal flow, and poloidal electron flow velocities, respectively.

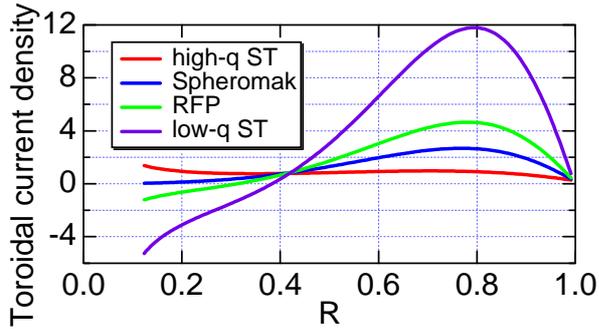

**FIGURE 4.** Radial profiles of toroidal current density on the midplane. The red, blue, green, and purple lines indicate the high-$q$ HD-ST ($C_{Hi1}=1.0$), the helicity-driven spheromak ($C_{Hi1}=9.2$), the helicity-driven RFP ($C_{Hi1}=15.0$), and the ultra low-$q$ HD-ST ($C_{Hi1}=28.0$).

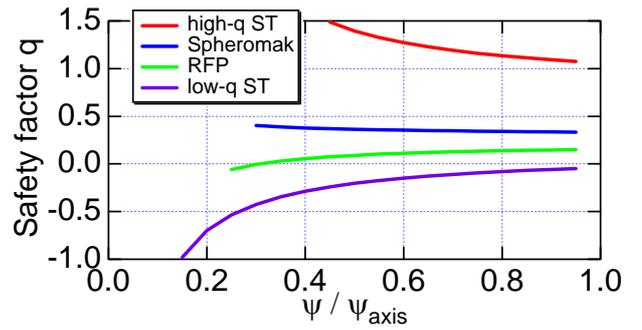

**FIGURE 5.** Safety factor $q$ as a function of the normalized poloidal flux function $\psi/\psi_{axis}$ for the same condition as Fig. 4. Here $\psi_{axis}$ is $\psi$ at the magnetic axis.

We show the safety factor $q$ as a function of the normalized poloidal flux function $\psi/\psi_{axis}$ in Fig. 5. As the effect of the ion flow becomes larger, the $q$-value comes down and reverses the sign at the inner edge region. Finally, it reverses the sign at the whole region, and becomes the ultra low-$q$ ($0<q<1$).

The polodal flux contours are shown in Fig. 6. All these flux surfaces have the open flux penetrating the electrodes, and form the helicity-driven configurations. This suggests the possibility of the current drive by coaxial helicity injection. As the effect of the ion flow becomes more significant, the amount of closed flux increases. The HD-STs have significantly lower $<\lambda>$ values than the helicity-driven spheromak and RFP. Note that the ultra low-$q$ HD-ST with diamagnetic $B_t$ and high-$\beta$ appears in the regime of $<\lambda>$ value lower than the lowest eigenvalue $\lambda_e$=9.29 m$^{-1}$. Therefore, it could be observed in the experiment.

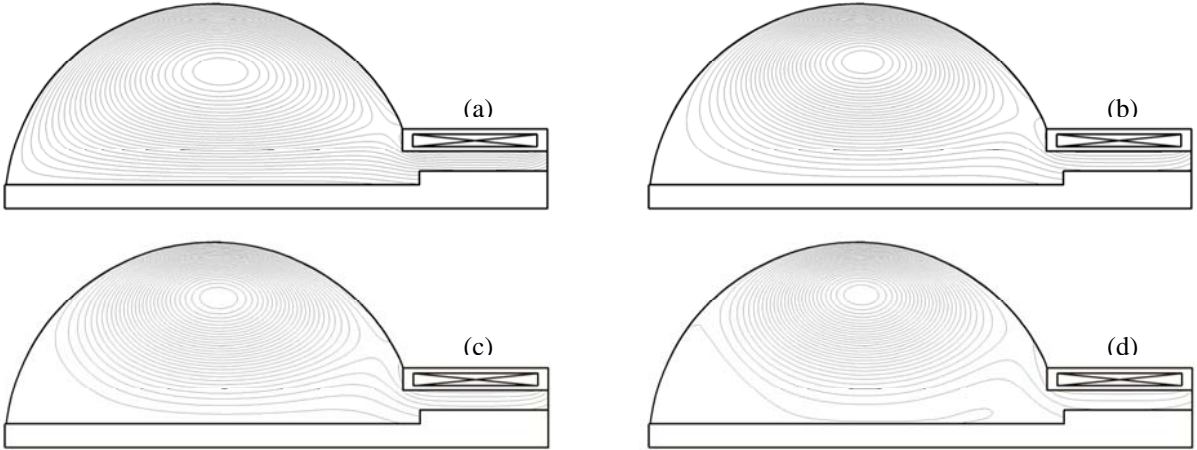

**FIGURE 6.** Poloidal flux contours for the same condition as Fig. 2.

We examine the dependence of the maximum value of the toroidal ion flow $u_{itmax}$ on $<\beta>$ and $f_{2F}$ in Fig. 7. As $u_{itmax}$ increases with the transition of the high-$q$ HD-ST to the helicity-driven RFP, $<\beta>$ increases due to the decrease in $B_{t.e}$. On the other hand, $<\beta>$ gradually decreases as $u_{itmax}$ increases from the helicity-driven RFP to the ultra low-$q$ HD-ST. It is also found from Fig. 7(b) that except for the region of negative $u_{itmax}$, all the values of $f_{2F}$ are larger than unity. In the region of slow ion flow, $f_{2F}$ has a sharply peaked value ($u_{itmax}$=0.032).

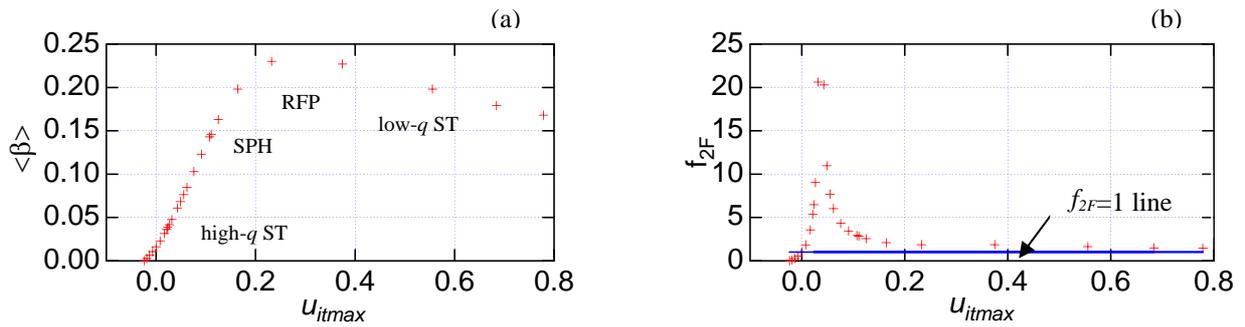

**FIGURE 7.** Dependence of the value of $u_{itmax}$ on $<\beta>$ and $f_{2F}$. Here $u_{itmax}$ represents the maximum (minimum) value of the toroidal ion flow velocity $u_{it}$ when $u_{it}$ is positive (negative).

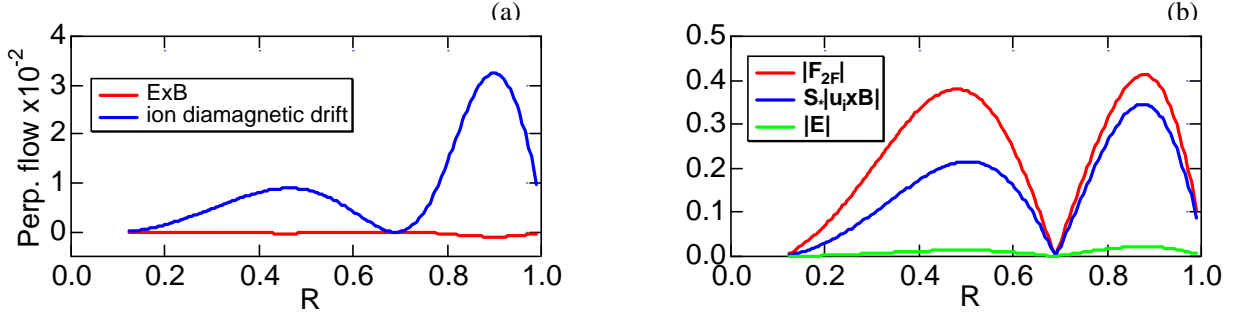

**FIGURE 8.** Radial profiles of fluid drifts, $|\mathbf{F}_{2F}|$, $|S_*\mathbf{u}_i \times \mathbf{B}|$, and $|\mathbf{E}|$ on the midplane for $u_{itmax}$=0.032. (a) $\mathbf{E} \times \mathbf{B}$ drift (red line) and ion diamagnetic drift (blue line), and (b) $|\mathbf{F}_{2F}|$ (red line), $|S_*\mathbf{u}_i \times \mathbf{B}|$ (blue), and $|\mathbf{E}|$ (green).

We consider why $f_{2F}$ has the sharply peaked value. Figure 8 shows the radial profiles of fluid drifts, $|\mathbf{F}_{2F}|$, $|S_*\mathbf{u}_i \times \mathbf{B}|$, and $|\mathbf{E}|$ on the midplane. In the case of $u_{itmax}$=0.032, it is found from Fig. 8(a) that the $\mathbf{E} \times \mathbf{B}$ drift is approximately zero, and the ion diamagnetic drift is dominant. Due to $|\mathbf{E}| \approx 0$, the balance is maintained by $S_*\mathbf{u}_i \times \mathbf{B}$ and $\mathbf{F}_{2F}$ as shown in Fig. 8(b). Therefore, $f_{2F}$ becomes significantly large.

## CONCLUSIONS

We have investigated the two-fluid effects on the MHD equilibrium configurations of the HIST HD-ST. Conclusions obtained in this paper are summarized as follows. 1) Equilibrium of the HD-ST based on the two-fluid model with flow in the region including the FC and CHS are numerically determined by using the finite difference and the boundary element methods. 2) The magnetic configurations change from the high-$q$ HD-ST ($q>1$) with paramagnetic toroidal field and low-$\beta$ ($<\beta> \approx 2$ %) through the helicity-driven spheromak and RFP to the ultra low-$q$ HD-ST ($0<q<1$) with diamagnetic toroidal field and high-$\beta$ ($<\beta> \approx 18$ %) as the external toroidal field at the inner edge regions decreases and reverses the sign. 3) In the ultra low-$q$ HD-ST, the toroidal field reverses the sign, but the poloidal field does not do it. Thus, it is different from the flipped ST observed in the experiment. Also, the ultra low-$q$ HD-ST appears in the regime of $<\lambda>$ value ($<\lambda>$= -0.673 m$^{-1}$) lower than the lowest eigenvalue $\lambda_e$ =9.29 m$^{-1}$. Therefore, it could be observed in the experiment. 4) The two-fluid effects are more significant in this equilibrium transition when the ion diamagnetic drift is dominant in the flowing two-fluid.

The fundamental properties of the HD-ST equilibrium based on the two-fluid model with flow outlined here are generally very available for predicting what equilibrium configuration is formed in the HD-ST experiment. There are further issues related to the equilibrium of the ultra low-$q$ HD-ST: 1) Can the generalized helicities conserve during this equilibrium transition? 2) How do we experimentally drive a flow of Alfven Mach number $M_A \approx 0.7$ for producing the ultra low-$q$ HD-ST? 3) Stability analysis of the flowing two-fluid equilibrium of the HD-ST is required.

## ACKNOLEDGMENTS

This work was partially supported by the Electric Technology Research Foundation of Chugoku.